\documentclass[aps,prl,showpacs]{revtex4}
\usepackage{epsfig,graphicx}

\input{tcilatex}

\begin{document}

\title{ Bell Bases Decomposition of a General N-qubit State
Teleportation through a Non-maximally Entangled Quantum Channel}
\author{Xiu-Lao Tian, Xiao-Qiang Xi}
\address{Department of Applied Mathematics and Physics, Xi'an Institute of Posts and Telecommunications, Xi'an 710061,
China}

\date{\today}

\begin{abstract}
We propose a method of the Bell bases decomposition to teleport an
arbitrary unknown N-qubit state through a nonmaximally entangled
quantum channel, and give the universal decomposition matrix of
N-qubit. Using the decomposition matrix, we can easily obtain the
collapsed state at the receiver site. The inverse matrix that the
decomposition matrix is just the transformation matrix that the
receiver can manipulate. The decomposition matrix is a function of
the parameters of the quantum channel. After defining the sub-matrix
of the quantum channel, we find that the decomposition matrix is a
tensor product of the sub-matrixes.
\end{abstract}

\pacs{03.67.Hk}

\maketitle

\section{1. Introduction}

Quantum entanglement plays an important role in the field of quantum
information processing, such as quantum computation, quantum
cryptography, quantum teleportation, dense coding and so on
\cite{Nielsen}. Since Bennett {\em et al.} \cite{Bennett1895}
presented a quantum teleportation scheme, there has been a great
development of quantum communication in experiments
\cite{BouwmeesterN390_575,ZhengPRA63_044302,
SolanoEPJD13_121,NielsenN396_52} and theory. The investigations of N
(N$\geq 2$) qubits teleportation have attracted much attention, such
as, two-qubit teleportation \cite{LeePAR64_014302} and probabilistic
teleportation \cite{GaoCPL20_2094}, teleportation of a special GHZ
state \cite{YangCPL16_628}, probabilistically teleport a
three-particle state via three pairs of entangled particles
\cite{FangPRA67_014305}. Besides the discrete-variable systems,
continuous-variable systems
\cite{BowenPRA67_032302,JohnsonPRA66_042326} are also studied in
teleportation.

   There are three key problems and a difficult problem
in the processing of teleportation. The key problems are: (1) the
sender (Alice) and the receiver (Bob) must share the quantum
channels; (2) Alice performs the Bell bases measurement on her
particles; (3) Bob performs a corresponding transformation to his
particles to reestablish the transmitted state with a certain
probability. The difficult problem is multi-qubit teleportation. It
is well known that the Bell bases measurement is equivalent to the
joint Bell bases acting on the joint system, if we can decompose the
joint system with Bell bases state, then we can easily obtain the
results of the Bell bases measurement, it is meaningful to find the
Bell bases decomposition of the joint system. In a practical
teleportation, due to the effect of the environment and the channel
noises, the quantum channel is not always in a maximally entangled
state (MES). Therefore, it is very important to study the quantum
teleportation of an unknown quantum state by using a nonmaximally
entangled channel (NMES) \cite{ShiPLA268_161,LiPRA61_34301}.

    Recently, several researchers investigated  N($\geq
2$)-qubit teleportation, but they only use a special nonmaximally
entangled quantum channel or teleport a special state. Here we use
GENERAL in order to different from that of Ref.
\cite{ZhengAPS52_2678,Rigolin71_032303,YeoPRL96_060502,ChenPRA74_032324},
Certainly, it is very complex and difficult to study a GENERAL
N($\geq 2$)-qubit teleportation through a GENERAL NONMAXIMALLY
(GGNM). In this paper, in order to solve this GGNM problems, we
propose a method of the Bell bases decomposition in teleportation
and give the universal decomposition matrix of N-qubit teleportation
and maybe  the tensor product of the sub-matrixes can open out some
characteristic of teleportation. Using decomposition matrix, we can easily obtain
some useful information about teleportation.

\section{2. Bell bases decompositon of a general three-qubit and four-qubit state teleportation}

\subsection{2.1 Three-qubit case}

Supposing Alice wants to send an unknown general three-qubit state
$|\varphi>_{1,2,3}$ to Bob,

\begin{eqnarray}
|\varphi>_{1,2,3}&=&\sum_{i=1}^{2^3}X_i|x_i^{1}x_i^{2}x_i^{3}> \nonumber \\
&=&X_1|000>+X_2|001>+X_3|010>+X_4|011>+X_5|100>+X_6|101>+X_7|110>+X_8|111>.
\end{eqnarray}

Using three general entangled pairs as the quantum channel, they are
 \begin{eqnarray}
|\varphi>_{4,5}&=&\sum_{j^1=1}^{4}Y_{j}^1|y_{j}^1y_{j}^{'1}>=Y_{1}^1|00>+Y_{2}^1|01>+Y_{3}^1|10>+Y_{4}^1|11>, \\
|\varphi>_{6,7}&=&\sum_{j^2=1}^{4}Y_{j}^2|y_{j}^2y_{j}^{'2}>=Y_{1}^2|00>+Y_{2}^2|01>+Y_{3}^2|10>+Y_{4}^2|11>, \\
|\varphi>_{8,9}&=&\sum_{j^3=1}^{4}Y_{j}^2|y_{j}^3y_{j}^{'3}>=Y_{1}^3|00>+Y_{2}^3|01>+Y_{3}^3|10>+Y_{4}^3|11>.
\end{eqnarray}

Where $j^{i}$($i=1,2,3$) denotes the $i$-th entangled pair of the
quantum channel, similarly later. The state of the joint system can
be written as

\begin{equation}
|\Psi_T>=|\varphi_{1,2,3}\otimes
\varphi_{4,5}\otimes\varphi_{6,7}\otimes\varphi_{8,9}>=
\sum_{i=1}^4\sum_{j^1=1}^4\sum_{j^2=1}^4\sum_{j^3=1}^4X_iY_j^{1}Y_j^{2}Y_j^{3}
|x_i^{1}x_i^{2}x_i^{3}
y_{j}^{1}y_{j}^{'1}y_{j}^{2}y_{j}^{'2}y_{j}^{3}y_{j}^{'3}>_{1-9},
\end{equation}
 where $X_iY_{j}^{1}Y_{j}^{2}Y_{j}^{3}$ is the coefficient,
  $|x_i^{1}x_i^2x_i^3>$ instead
 of the micro state of three particle and $|y_{j}^{i}y_{j}^{'i}>$ ($i=1,2,3$) instead of
 the micro state of two particle. The
 value of the indexes in different qubits can be seen from
 Eq.(1)-(4), they are
 $x_{1,2,3,4}^{1}=x_{1,2,5,6}^{2}=x_{1,3,5,7}^{3}=0$,
 $x_{5,6,7,8}^{1}=x_{3,4,7,9}^{2}=x_{2,4,6,8}^{3}=1$,
 $y_{1,2}^{1}=y_{1,3}^{'1}=y_{1,2}^{2}=y_{1,3}^{'2}=y_{1,2}^{3}=y_{1,3}^{'3}=0$,
  $y_{3,4}^{1}=y_{2,4}^{'1}=y_{3,4}^{2}=y_{2,4}^{'2}=y_{3,4}^{3}=y_{2,4}^{'3}=1$.
Let particles 1,2,3,5,7,9 belong to Alice and 4,6,8 belong to Bob,
 Alice can measure $|\Psi_T>$ with joint Bell bases
 $|\varphi_{1,5}^{\alpha_1}\varphi_{2,7}^{\alpha_{2}}
 \varphi_{3,9}^{\alpha_{3}}>$, where $\varphi_{i,j}^{k}$
 is the Bell bases $|\varphi_{i,j}^{1}>=(|00>+|11>)/\sqrt{2},
 |\varphi_{i,j}^{2}>=(|00>-|11>)/\sqrt{2},|\varphi_{i,j}^{3}>
 (|01>+|10>)/\sqrt{2},
 |\varphi_{i,j}^{4}>=(|01>-|10>)/\sqrt{2}$.

So we can rearrange the qubits in $|\Psi_T>$ by the order of
1-5-2-7-3-9-4-6-8 and rewrite $|\Psi_T>$ as
\begin{equation}
|\Psi_T>=\sum_{i=1}^4\sum_{j^1=1}^4\sum_{j^2=1}^4\sum_{j^3=1}^4X_iY_j^{1}Y_j^{2}Y_j^{3}
|(x_i^{1}y_{j}^{'1}x_i^{2}y_{j}^{'2}x_i^{3}y_{j}^{'3})_{1,5,2,7,3,9}(y_{j}^{1}
y_{j}^{2}y_{j}^{3})_{4,6,8}>.
\end{equation}

Under the bases of three qubit
($|000>,|001>,|010>,|011>,|100>,|101>,|110>,|111>$), the matrix
expression of $|\Psi_T>$ by the Bell bases decomposition of
$|\varphi_{1,5}^{\alpha_{1}}\varphi_{2,7}^{\alpha_{2}}\varphi_{3,9}^{\alpha_{3}}>$
is
\begin{equation}
|\Psi_T>=(1/\sqrt{2})^3\sum_{\alpha_{1}=1}^4\sum_{\alpha_{2}=1}^4
\sum_{\alpha_{3}=1}^4\varphi_{1,5}^{\alpha_{1}}\varphi_{2,7}^{\alpha_{2}}\varphi_{3,9}^{\alpha_{3}}
\sigma_{4,6,8}^{\alpha_{1}\alpha_{2}\alpha_{3}}|\varphi_{4,6,8}>,
\end{equation}
where $|\varphi_{4,6,8}>=(X_1,X_2,\dots,X_8)_{4,6,8}^T$,
$\sigma_{4,6,8}^{\alpha_{1}\alpha_{2}\alpha_{3}}$ is called the
decomposition matrix. Alice measures $|\Psi_T>$ with the joint Bell
bases of
$|\varphi_{1,5}^{\alpha_{1}}\varphi_{2,7}^{\alpha_{2}}\varphi_{3,9}^{\alpha_{3}}>$
is equivalent to the action of
$|\varphi_{1,5}^{\alpha_{1}}\varphi_{2,7}^{\alpha_{2}}\varphi_{3,9}^{\alpha_{3}}>$
on $|\psi_T>$, and Bob gets the possible collapsed quantum state
$(1/\sqrt{2})^3\sigma_{4,6,8}^{\alpha_{1}\alpha_{2}\alpha_{3}}|\varphi_{4,6,8}>$,
i.e.
\begin{equation}
<\varphi_{1,5}^{\alpha_{1}}\varphi_{2,7}^{\alpha_{2}}\varphi_{3,9}^{\alpha_{3}}|\Psi_T>=
(1/\sqrt{2})^3\sigma_{4,6,8}^{\alpha_{1}\alpha_{2}\alpha_{3}}|\varphi_{4,6,8}>.
\end{equation}

We can calculate the decomposition matrixes
$\sigma_{4,6,8}^{\alpha_{1}\alpha_{2}\alpha_{3}}|\varphi_{4,6,8}>$
through Eq.(8) and induce its general expressions,
\begin{equation}
|\varphi_{1,5}^{1}\varphi_{2,7}^{1}\varphi_{3,9}^{1}>=(1/\sqrt{2})^3
(|00>+|11>)_{1,5}(|00>+|11>)_{2,7}(|00>+|11>)_{3,9}.
\end{equation}

From the value of the indexes in different qubits, we can get the
micro states in Eq.(9), see Table 1.

\centerline{\footnotesize Table 1. The states under general bases
and the joint Bell bases
$|\varphi_{1,5}^{1}\varphi_{2,7}^{1}\varphi_{3,9}^{1}>$.}

\begin{center}
{\footnotesize
\begin{tabular}{c|c||c|c}
\hline state & state under joint Bell bases & state& state under
joint Bell bases\cr \hline

$|000000>$&
$\sum_{j^1=1,3}\sum_{j^2=1,3}\sum_{j^3=1,3}X_1Y_{j}^{1}Y_{j}^{2}Y_{j}^{3}
|x_1^{1}y_{j}^{'1}x_1^{2}y_{j}^{'2}x_1^{3}y_{j}^{'3}>$ &$|000011>$&
$\sum_{j^1=1,3}\sum_{j^2=1,3}\sum_{j^3=2,4}X_2^{2}Y_{j}^{1}Y_{j}^{2}Y_{j}^{3}
|x_2^{1}y_{j}^{'1}x_2^{2}y_{j}^{'2}x_2^{3}y_{j}^{'3}>$ \cr \hline

$|001100>$&
$\sum_{j^1=1,3}\sum_{j^2=2,4}\sum_{j^3=1,3}X_3Y_{j}^{1}Y_{j}^{2}Y_{j}^{3}
|x_3^{1}y_{j}^{'1}x_3^{2}y_{j}^{'2}x_3^{3}y_{j}^{'3}>$ &$|001111>$&
$\sum_{j^1=1,3}\sum_{j^2=2,4}\sum_{j^3=2,4}X_4Y_{j}^{1}Y_{j}^{2}Y_{j}^{3}
|x_4^{1}y_{j}^{'1}x_4^{2}y_{j}^{'2}x_4^{3}y_{j}^{'3}>$ \cr \hline

$|110000>$&
$\sum_{j^1=2,4}\sum_{j^2=1,3}\sum_{j^3=1,3}X_5Y_{j}^{1}Y_{j}^{2}Y_{j}^{3}
|x_5^{1}y_{j}^{'1}x_5^{2}y_{j}^{'2}x_5^{3}y_{j}^{'3}>$ &$|110011>$&
$\sum_{j^1=2,4}\sum_{j^2=1,3}\sum_{j^3=2,4}X_6Y_{j}^{1}Y_{j}^{2}Y_{j}^{3}
|x_6^{1}y_{j}^{'1}x_6^{2}y_{j}^{'2}x_6^{3}y_{j}^{'3}>$ \cr \hline

$|111100>$&
$\sum_{j^1=2,4}\sum_{j^2=2,4}\sum_{j^3=1,3}X_7Y_{j}^{1}Y_{j}^{2}Y_{j}^{3}
|x_7^{1}y_{j}^{'1}x_7^{2}y_{j}^{'2}x_7^{3}y_{j}^{'3}>$ &$|111111>$&
$\sum_{j^1=2,4}\sum_{j^2=2,4}\sum_{j^3=2,4}X_8Y_{j}^{1}Y_{j}^{2}Y_{j}^{3}
|x_8^{1}y_{j}^{'1}x_8^{2}y_{j}^{'2}x_8^{3}y_{j}^{'3}>$ \cr \hline

\end{tabular}}
\end{center}

$\sigma_{4,6,8}^{111}|\varphi_{4,6,8}>$ can be written as
\begin{eqnarray}
\sigma_{4,6,8}^{111}|\varphi_{4,6,8}>&=&[X_1
\sum_{j^1=1,3}\sum_{j^2=1,3}\sum_{j^3=1,3}+X_2
\sum_{j^1=1,3}\sum_{j^2=1,3}\sum_{j^3=2,4}+X_3
\sum_{j^1=1,3}\sum_{j^2=2,4}\sum_{j^3=1,3}\nonumber \\
 &+&X_4
\sum_{j^1=1,3}\sum_{j^2=2,4}\sum_{j^3=2,4}+X_5
\sum_{j^1=2,4}\sum_{j^2=1,3}\sum_{j^3=1,3}+X_6
\sum_{j^1=2,4}\sum_{j^2=1,3}\sum_{j^3=2,4}\nonumber \\
 &+&X_7
\sum_{j^1=2,4}\sum_{j^2=2,4}\sum_{j^3=1,3}+X_8
\sum_{j^1=2,4}\sum_{j^2=2,4}\sum_{j^3=2,4}]Y_{j}^{1}Y_{j}^{2}Y_{j}^{3}
|y_{j}^{1}y_{j}^{2}y_{j}^{3}>_{4,6,8}.
\end{eqnarray}

Through tedious calculation and using $Y_j^{1}Y_j^{2}Y_j^{3}$
instead of $Y_j^{1}Y_j^{2}Y_j^{3}|y_j^{1}y_j^{2}y_j^{3}>$ we get
\begin{eqnarray}
\sigma_{4,6,8}^{111}|\varphi_{4,6,8}>&=&X_1(Y_{1}^{1}+Y_{3}^{1})
(Y_{1}^{2}+Y_{3}^{2})(Y_{1}^{3}+Y_{3}^{3})+
X_2(Y_{1}^{1}+Y_{3}^{1})(Y_{1}^{2}+Y_{3}^{2})(Y_{2}^{3}+Y_{4}^{3}) \nonumber \\
&+&X_3(Y_{1}^{1}+Y_{3}^{1})(Y_{2}^{2}+Y_{4}^{2})(Y_{1}^{3}+Y_{3}^{3})
+X_4(Y_{1}^{1}+Y_{3}^{1})(Y_{2}^{2}+Y_{4}^{2})(Y_{2}^{3}+Y_{4}^{3})
\nonumber \\
&+&X_5(Y_{2}^{1}+Y_{4}^{1})(Y_{1}^{2}+Y_{3}^{2})(Y_{1}^{3}+Y_{3}^{3})
+X_6(Y_{2}^{1}+Y_{4}^{1})(Y_{1}^{2}+Y_{3}^{2})(Y_{2}^{3}+Y_{4}^{3})
\nonumber \\
&+&X_7(Y_{2}^{1}+Y_{4}^{1})(Y_{2}^{2}+Y_{4}^{2})(Y_{1}^{3}+Y_{3}^{3})
+X_8(Y_{2}^{1}+Y_{4}^{1})(Y_{2}^{2}+Y_{4}^{2})(Y_{2}^{3}+Y_{4}^{3}).
\end{eqnarray}

In the bases of $(X_1,X_2,\dots,X_8)_{4,6,8}^T$,
$\sigma_{4,6,8}^{111}$ can be written as the tensor product:

\begin{equation}
\sigma_{4,6,8}^{111}=
\left(\begin{array}{cc} Y_{1}^{1}& Y_{2}^{1}\\
 Y_{3}^{1}& Y_{4}^{1}\end{array}\right)\otimes
 \left(\begin{array}{cc} Y_{1}^{2}& Y_{2}^{2}\\
 Y_{3}^{2}& Y_{4}^{2}\end{array}\right)\otimes
 \left(\begin{array}{cc} Y_{1}^{3}& Y_{2}^{3}\\
 Y_{3}^{3}& Y_{4}^{3}\end{array}\right).
\end{equation}

Similarly, we can obtain
\begin{equation}
\sigma_{4,6,8}^{234}=
\left(\begin{array}{cc} Y_{1}^{1}& -Y_{2}^{1}\\
 Y_{3}^{1}& -Y_{4}^{1}\end{array}\right)\otimes
 \left(\begin{array}{cc} Y_{2}^{2}& Y_{1}^{2}\\
 Y_{4}^{2}& Y_{3}^{2}\end{array}\right)\otimes
 \left(\begin{array}{cc} Y_{2}^{3}& -Y_{1}^{3}\\
 Y_{4}^{3}& -Y_{3}^{3}\end{array}\right).
\end{equation}

Through analysis and validation we find that
$\sigma_{4,6,8}^{\alpha_{1}\alpha_{2}\alpha_{3}}$ has an general
expression
\begin{equation}
\sigma_{4,6,8}^{\alpha_{1}\alpha_{2}\alpha_{3}}=\sigma_4^{\alpha_{1}}\otimes
\sigma_6^{\alpha_{2}}\otimes\sigma_8^{\alpha_{3}},
\end{equation}
where $\alpha_{1},\alpha_{2},\alpha_{3}=1,2,3,4$ and
$\sigma_4^{\alpha_{1}},\sigma_6^{\alpha_{2}},\sigma_8^{\alpha_{3}}$
are called the sub-matrixes of the decomposition matrix
$\sigma_{4,6,8}^{\alpha_{1}\alpha_{2}\alpha_{3}}$, the expression of
them are shown in Table 2.

\centerline{\footnotesize Table 2. The sub-matrixes of the
three-qubit decomposition matrix, $i=1,2,3$ instead of the $i$-th
entangled pair.}

\begin{center}
\begin{tabular}{c|c|c|c}
\hline   $\sigma^1$ & $\sigma^2$&$\sigma^3$ & $\sigma^4$\cr \hline
 $\left( \begin{array}{cc}Y_{1}^{i}& Y_{2}^{i}\\ Y_{3}^{i}&
 Y_{4}^{i}\end{array}\right)$
 &  $\left( \begin{array}{cc}Y_{1}^{i}& -Y_{2}^{i}\\ Y_{3}^{i}&
 -Y_{4}^{i}\end{array}\right)$
 &  $\left( \begin{array}{cc}Y_{2}^{i}& Y_{1}^{i}\\ Y_{4}^{i}&
 Y_{3}^{i}\end{array}\right)$
  &  $\left( \begin{array}{cc}Y_{2}^{i}& -Y_{1}^{i}\\ Y_{4}^{i}&
   -Y_{3}^{i}\end{array}\right)$\cr \hline
\end{tabular}
\end{center}

$\sigma_{4,6,8}^{\alpha_{1}\alpha_{2}\alpha_{3}}$ is a function of
the parameters of the quantum channel$'$ , different decomposition
matrix corresponds to the different quantum channels.

Using the method of tensor product, we find that the sub-matrixes
have a universal expression
\begin{equation}
\sigma_{i}^{\mu}= \left(\begin{array}{cc} T_{\mu1}^{i}Y_{1}^{i}
+T_{\mu2}^{i}Y_{2}^{i} & T_{\mu3}^{i}Y_{1}^{i}
+T_{\mu4}^{i}Y_{2}^{i} \\
 T_{\mu1}^{i}Y_{3}^{i}+T_{\mu2}^{i}Y_{4}^{i} &
 T_{\mu3}^{i}Y_{3}^{i}
 +T_{\mu4}^{i}Y_{4}^{i} \end{array}\right),
\end{equation}
where $\mu=1,2,3,4$, $i=1,2,3$ and $T_{\mu k}(k=1,2,3,4)$ is the
elements of the transformation matrix between the Bell bases and the
general bases of two-qubit ($|00>,|01>,|10>,|11>$)
\begin{equation}
\left(\begin{array}{c}\varphi_{i,j}^1\\ \varphi_{i,j}^2
\\ \varphi_{i,j}^3 \\ \varphi_{i,j}^4
 \end{array} \right)=\frac{1}{\sqrt{2}}T_{ij}
 \left(\begin{array}{c}00 \\ 01
\\ 10 \\ 11
 \end{array} \right)_{ij}=\frac{1}{\sqrt{2}}
\left(\begin{array}{cccc} 1 & 0 & 0 & 1\\
 1 & 0 & 0 & -1\\
 0 & 1 & 1 & 0\\
 0 & 1 & -1 & 0
 \end{array}\right)_{ij}\left(\begin{array}{c}00 \\ 01
\\ 10 \\ 11
 \end{array} \right)_{ij}.
\end{equation}

\subsection{2.2 Four-qubit case}

Supposing Alice wants to send an unknown general four-qubit state
$|\varphi>_{1,2,3,4}$ to Bob,

\begin{eqnarray}
|\varphi>_{1,2,3,4}&=&\sum_{i=1}^{2^4}X_i|x_i^{1}x_i^{2}x_i^{3}x_i^{4}> \nonumber \\
&=&X_1|0000>+X_2|0001>+X_3|0010>+X_4|0011>+X_5|0100>+X_6|0101> \nonumber \\
& &
+X_7|0110>+X_8|0111>+X_9|1000>+X_{10}|1001>+X_{11}|1010>+X_{12}|1011> \nonumber \\
& & +X_{13}|1100>+X_{14}|1101>+X_{15}|1110>+X_{16}|1111>.
\end{eqnarray}

Using four entangled pairs as the quantum channel, they are
 \begin{eqnarray}
|\varphi>_{5,6}&=&\sum_{j^1=1}^{4}Y_j^{1}|y_j^{1}y_j^{'1}>_{5,6}
=Y_{1}^{1}|00>+Y_{2}^{1}|01>+Y_{3}^{1}|10>+Y_{4}^{1}|11>, \\
|\varphi>_{7,8}&=&\sum_{j^2=1}^{4}Y_j^{2}|y_j^{1}y_j^{'2}>_{7,8}
=Y_{1}^{2}|00>+Y_{2}^{2}|01>+Y_{3}^{2}|10>+Y_{4}^{2}|11>, \\
|\varphi>_{9,10}&=&\sum_{j^3=1}^{4}Y_j^{3}|y_j^{1}y_j^{'3}>_{9,10}
=Y_{1}^{3}|00>+Y_{2}^{3}|01>+Y_{3}^{3}|10>+Y_{4}^{3}|11>,\\
|\varphi>_{11,12}&=&\sum_{j^4=1}^{4}Y_j^{4}|y_j^{1}y_j^{'4}>_{11,12}
=Y_{1}^{4}|00>+Y_{2}^{4}|01>+Y_{3}^{4}|10>+Y_{4}^{4}|11>.
\end{eqnarray}

The state of the joint system can be written as

\begin{equation}
|\Psi_T>=|\varphi_{1,2,3,4}\otimes
\varphi_{5,6}\otimes\varphi_{7,8}\otimes\varphi_{9,10}\otimes\varphi_{11,12}>=
\sum_i^4\sum_{j^1}^4\sum_{j^2}^4\sum_{j^3}^4
\sum_{j^4}^4X_iY_j^{1}Y_j^{2}Y_j^{3}Y_j^{4}|x_i^{1}x_i^{2}x_i^{3}x_i^{4}
y_j^{1}y_j^{'1}y_j^{2}y_j^{'2}y_j^{3}y_j^{'3}y_j^{4}y_j^{'4}>.
\end{equation}

Let particles 1,2,3,4,5,7,9,11 belong to Alice and 6,8,10,12 belong
to Bob,
 then Alice can measure $|\Psi_T>$ with joint Bell bases
 $|\varphi_{1,5}^{\alpha_{1}}\varphi_{2,7}^{\alpha_{2}}
 \varphi_{3,9}^{\alpha_{3}}\varphi_{4,11}^{\alpha_{4}}>$.

So we can rearrange the qubits in $|\Psi_T>$ by the order of
1-5-2-7-3-9-4-11-6-8-10-12 and rewrite $|\Psi_T>$ as
\begin{equation}
|\Psi_T>=\sum_{i=1}^4\sum_{j^1=1}^4\sum_{j^2=1}^4\sum_{j^3=1}^4
\sum_{j^4=1}^4X_iY_j^{1}Y_j^{2}Y_j^{3}Y_j^{4}
|(x_i^{1}y_j^{1}x_i^{2}y_j^{2}x_i^{3}y_j^{3}x_i^{4}
y_j^{4})_{1,5,2,7,3,9,4,11}(y_j^{'1}y_j^{'2}y_j^{'3}y_j^{'4})_{6,8,10,12}>.
\end{equation}

Under the bases of four qubit, the matrix expression of $|\Psi_T>$
by the Bell bases decomposition of
$|\varphi_{1,5}^{\alpha_{1}}\varphi_{2,7}^{\alpha_{2}}
\varphi_{3,9}^{\alpha_{3}}\varphi_{4,11}^{\alpha_{4}}>$ is
\begin{equation}
|\Psi_T>=(1/\sqrt{2})^4\sum_{\alpha_{1}=1}^4\sum_{\alpha_{2}=1}^4
\sum_{\alpha_{3}=1}^4\sum_{\alpha_{4}=1}^4
\varphi_{1,5}^{\alpha_{1}}\varphi_{2,7}^{\alpha_{2}}\varphi_{3,9}^{\alpha_{3}}
\varphi_{4,11}^{\alpha_{4}}\sigma_{6,8,10,12}^{\alpha_{1}
\alpha_{2}\alpha_{3}\alpha_{4}}|\varphi_{6,8,10,12}>,
\end{equation}
where $|\varphi>_{6,8,10,12}=(X_1,X_2,\dots,X_{16})_{6,8,10,12}^T$,
$\sigma_{6,8,10,12}^{\alpha_{1}\alpha_{2}\alpha_{3}}$ is the
decomposition matrix of four-qubit state teleportation. Alice
measures $|\Psi_T>$ with the joint Bell bases of
$|\varphi_{1,5}^{\alpha_{1}}\varphi_{2,7}^{\alpha_{2}}
\varphi_{3,9}^{\alpha_{3}}\varphi_{4,11}^{\alpha_{4}}>$, the
collapsed quantum state that Bob can obtain is
$|(1/\sqrt{2})^4\sigma_{6,8,10,12}^{\alpha_{1}\alpha_{2}
\alpha_{3}\alpha_{4}}|\varphi_{6,8,10,12}>$, i.e.
\begin{equation}
<\varphi_{1,5}^{\alpha_{1}}\varphi_{2,7}^{\alpha_{2}}
\varphi_{3,9}^{\alpha_{3}}\varphi_{4,11}^{\alpha_{3}}|\Psi_T>=
(1/\sqrt{2})^4\sigma_{6,8,10,12}^{\alpha_{1}\alpha_{2}
\alpha_{3}\alpha_{4}}|\varphi_{6,8,10,12}>.
\end{equation}

The general expression of
$\sigma_{6,8,10,12}^{\alpha_{1}\alpha_{2}\alpha_{3}\alpha_{4}}$ is
\begin{equation}
\sigma_{6,8,10,12}^{\alpha_{1}\alpha_{2}\alpha_{3}\alpha_{4}}=\sigma_6^{\alpha_{1}}\otimes
\sigma_8^{\alpha_{2}}\otimes\sigma_{10}^{\alpha_{3}}\otimes\sigma_{12}^{\alpha_{4}},
\end{equation}
where $\alpha_{1},\alpha_{2},\alpha_{3},\alpha_{4}=1,2,3,4$ and
$\sigma_6^{\alpha_{1}},\ \sigma_8^{\alpha_{2}},\sigma_{10}^{\alpha_{3}},
\sigma_{12}^{\alpha_{4}}$
are the sub-matrixes,the expression of them are shown in Table 3.

\vspace{0.6cm}
\centerline{\footnotesize Table 3. The sub-matrixes of
the four-qubit decomposition matrix, $i=1,2,3,4$ instead of the
$i$-th entangled pair.}

\begin{center}
\begin{tabular}{c|c|c|c}
\hline    $\sigma^1$ & $\sigma^2$& $\sigma^3$ & $\sigma^4$ \cr
\hline
  $\left( \begin{array}{cc}Y_1^{i}& Y_3^{i}\\ Y_2^{i}&
 Y_4^{i}\end{array}\right)$
 &  $\left( \begin{array}{cc}Y_1^{i}& -Y_3^{i}\\ Y_2^{i}&
 -Y_4^{i}\end{array}\right)$
 &  $\left( \begin{array}{cc}Y_3^{i}& Y_1^{i}\\ Y_4^{i}&
 Y_2^{i}\end{array}\right)$
  &  $\left( \begin{array}{cc}Y_3^{i}& -Y_1^{i}\\ Y_4^{i}& -Y_2^{i}\end{array}\right)$\cr \hline
 \end{tabular}
\end{center}

$\sigma_{6,8,10,12}^{\alpha_{1}\alpha_{2}\alpha_{3}\alpha_{4}}$ is
the function of the parameters of the quantum channel, different
quantum channel corresponds to differences decomposition matrix.
There have been some different between Table 3 and Table 2, because
we have chosen different Bell bases in three qubit and four qubit
cases.

\section{3. Bell bases decomposition of a general N-qubit state teleportation}

Supposing Alice wants to send an unknown N-qubit state
$|\varphi>_{1,2,\dots,N}$ to Bob,

\begin{equation}
|\varphi>_{1,2,\dots,N}=\sum_{i=1}^{2^N}X_i|x_i^{1}x_i^{2}\dots
x_i^{N}>.
\end{equation}

Using N entangled pairs as the quantum channel, they are
\begin{equation}
|\varphi>_{N+(2n-1),N+2n}=\sum_{j=1}^{4}Y_j^n|y_j^ny_j^{'n}>
=Y_{1}^n|00>+Y_{2}^n|01>+Y_{3}^n|10>+Y_{4}^n|11>,
\end{equation}
where $n=1,2,\dots,N$.

The state of the joint system can be written as

\begin{eqnarray}
|\Psi_T>&=&\varphi_{1,2,\dots,N}\otimes
\varphi_{N+1,N+2}\otimes\dots\otimes\varphi_{N+(2N-1),N+2N}
\nonumber \\
&=&
\sum_i^4\underbrace{\sum_{j^1}^4\dots\sum_{j^N}^4}_{N}X_iY_j^1Y_j^2\dots
Y_j^N|x_i^{1}x_i^{2}\dots x_i^{N} y_j^{1}y_j^{'1}\dots
y_j^{N}y_j^{'N}>_{1-3N},
\end{eqnarray}

Let particle $1,2,\dots,N,N+1,N+3,\dots,N+(2N-1)$ belong to Alice
and $N+2,N+4,\dots,N+2N$ belong to Bob,
 then Alice can measure $|\Psi_T>$ with Bell bases
 $|\varphi_{1,N+1}^{\alpha_1}\varphi_{2,N+3}^{\alpha_2}\dots\varphi_{N,N+(2N-1)}^{\alpha_N}>$
 $(\alpha_{1,2,\dots,N}=1,2,3,4)$,
  where $\varphi_{i,j}^{k}$
 is the Bell bases $|\varphi_{i,j}^{1}>=(|00>+|11>)/\sqrt{2},
 |\varphi_{i,j}^{2}>=(|00>-|11>)/\sqrt{2},|\varphi_{i,j}^{3}>=(|01>+|10>)/\sqrt{2},
 |\varphi_{i,j}^{4}>=(|01>-|10>)/\sqrt{2}$.

We can rearrange the order of the particles as
$1-(N+1)-2-(N+3)-\dots-N-(N+2N-1)-(N+2)-(N+4)-\dots-(N+2N)$ and
rewrite $|\Psi_T>$ as
\begin{equation}
|\Psi_T> =
\sum_i^4\underbrace{\sum_{j^1}^4\dots\sum_{j^N}^4}_{N}X_iY_j^1Y_j^2\dots
Y_j^N|(x_i^{1}y_j^{1}x_i^{2}y_j^{2}\dots x_i^{N}y_j^{N})
(y_j^{'1}\dots y_j^{'N})>_{1-3N}.
\end{equation}

In the bases of N qubit, the matrix expression of $|\Psi_T>$ by the
Bell bases decomposition of
$|\varphi_{1,N+1}^{\alpha_1}\varphi_{2,N+3}^{\alpha_2}\dots
\varphi_{N,N+(2N-1)}^{\alpha_N}>$
is
\begin{equation}
|\Psi_T>=(1/\sqrt{2})^N\sum_{\alpha_1=1}^4\sum_{\alpha_2=1}^4\dots
\sum_{\alpha_N=1}^4
\varphi_{1,N+1}^{\alpha_1}\varphi_{2,N+3}^{\alpha_2}\dots
\varphi_{N,N+(2N-1)}^{\alpha_N}
\sigma_{N+2,N+4,\dots,N+2N}^{\alpha_1\dots
\alpha_N}|\varphi_{N+2,N+4,\dots,N+2N}>
\end{equation}
where
$|\varphi_{N+2,N+4,\dots,N+2N}>=(X_1,X_2,\dots,X_{N})_{N+2,N+4,\dots,N+2N}^T$,
and $\sigma_{N+2,N+4,\dots,N+2N}^{\alpha_1\dots\alpha_N}$  is the
decomposition matrix. Alice measure $|\Psi_T>$ with the Bell bases
of $|\varphi_{1,N+1}^{\alpha_1}\varphi_{2,N+3}^{\alpha_2}
\dots\varphi_{N,N+(2N-1)}^{\alpha_N}>$, the collapsed quantum state
that Bob can obtain is
$(1/\sqrt{2})^N\sigma_{N+2,N+4,\dots,N+2N}^{\alpha_1\dots
\alpha_N}\varphi_{N+2,N+4,\dots,N+2N}$, i.e.
\begin{equation}
<\varphi_{1,N+1}^{\alpha_1}\varphi_{2,N+3}^{\alpha_2}
\dots\varphi_{N,N+(2N-1)}^{\alpha_N}|\Psi_T>=
(1/\sqrt{2})^N\sigma_{N+2,N+4,\dots,N+2N}^{\alpha_1\dots\alpha_N}
|\varphi_{N+2,N+4,\dots,N+2N}>.
\end{equation}

$\sigma_{N+2,N+4,\dots,N+2N}^{\alpha_1\dots \alpha_N}$ has the form
of
\begin{equation}
\sigma_{N+2,N+4,\dots,N+2N}^{\alpha_1\dots
\alpha_N}=\sigma_{N+2}^{\alpha_1}\otimes
\sigma_{N+4}^{\alpha_2}\otimes\dots\otimes\sigma_{N+2N}^{\alpha_N},
\end{equation}

Using the method of tensor product, we find that the sub-matrixes
have a simple universal expression
\begin{equation}
\sigma_{i}^{\mu}= \left(\begin{array}{cc} T_{\mu1}^{i}Y_{1}^{i}
+T_{\mu2}^{i}Y_{3}^{i} & T_{\mu3}^{i}Y_{1}^{i}
+T_{\mu4}^{i}Y_{3}^{i} \\
 T_{\mu1}^{i}Y_{2}^{i}+T_{\mu2}^{i}Y_{4}^{i} &
 T_{\mu3}^{i}Y_{2}^{i}
 +T_{\mu4}^{i}Y_{4}^{i} \end{array}\right),
\end{equation}
where $\mu=1,2,3,4$, $i=1,2,\dots,N$, the odd N, the even N and the
choice of Bell bases will affect the expression of
$\sigma_{i}^{\mu}$.

\section{4. The applications of the decomposition matrix}
This method of the Bell bases decomposition for arbitrary qubit
teleportation is universal.for example , Supposing Alice wants to
send an unknown general single-qubit state $|\varphi>_{1}$ to Bob,

\begin{equation}
|\varphi>_{1}=\sum_{i=1}^{2}X_i|x^i>=X_1|0>+X_2|1>.
\end{equation}
 the quantum channel is
 \begin{eqnarray}
|\varphi>_{2,3,}&=&\sum_{j^1=1}^{4}Y_{j}^1|y_{j}^1y_{j}^{'1}>=Y_{1}^1|00>+Y_{2}^1|01>+Y_{3}^1|10>+Y_{4}^1|11>
\end{eqnarray}
the joint system is
\begin{eqnarray}
|\Psi_T>_{1,2,3,}=\varphi_1\otimes\varphi_{2,3}
\end{eqnarray}
using Bell bases $|\varphi_{1,2}^{\alpha}>$ decompositon for the
joint system state
\begin{equation}
|\Psi_T>=(1/\sqrt{2})\sum_{\alpha=1}^4\varphi_{1,2}^{\alpha}\sigma_3^{\alpha}\varphi_3
\end{equation}
we get four decomposition matrixs.  $ {\sigma_2^1=
\left(\begin{array}{cc}
Y_{1}&Y_{3}\\
Y_{2}&Y_{4}\\
\end{array}\right)}$,
${\sigma_2^2= \left(\begin{array}{cc}
Y_{1}&-Y_{3}\\
Y_{2}&-Y_{4}\\
\end{array}\right)}$,
${\sigma_2^3= \left(\begin{array}{cc}
 Y_{3}&Y_{1}\\
Y_{4}&Y_{2}\\
\end{array}\right)}$,
${\sigma_2^4= \left(\begin{array}{cc}
 Y_{3}&-Y_{1}\\
Y_{4}&-Y_{2}\\
\end{array}\right)}$ .
The collapsed state that Bob can receive is
$(1/\sqrt{2})\sigma_{2,}^{\alpha}$, as long as there exists the
inverse matrix of $\sigma_{2}^{\alpha}$, Bob can obtain the
transported state. When
$Y_2^{1}=Y_3^{1}=Y_2^{2}=Y_3^{2}=Y_2^{3}=Y_3^{3}=0$, we get the
  same results as in Ref.\cite{LiPRA61_34301}.

 Furthermore , we consider  three-qubit case as the example.
 The collapsed state that Bob can receive is
$(1/\sqrt{2})\sigma_{4,6,8}^{\alpha_{1}\alpha_{2}\alpha_{3}}|\varphi_{4,6,8}>$,
as long as there exists the inverse matrix of
$\sigma_{4,6,8}^{\alpha_{1}\alpha_{2}\alpha_{3}}$, Bob can obtain
the transported state.

\begin{equation}
(\sigma_{4,6,8}^{\alpha_{1}\alpha_{2}\alpha_{3}})^{-1}=
(\sigma_4^{\alpha_{1}})^{-1}\otimes
(\sigma_6^{\alpha_{2}})^{-1}\otimes(\sigma_8^{\alpha_{3}})^{-1},
\end{equation}
for the quantum channel with parameters
$(Y_1^{i},Y_2^{i},Y_3^{i},Y_4^{i})$, the inverse matrix of the
sub-matrixes are {\footnotesize
\begin{eqnarray}
(\sigma_i^{1})^{-1}&=&\frac{1}{Y_1^{i}Y_4^{i}-Y_2^{i}Y_3^{i}}
\left( \begin{array}{cc}Y_4^{i}& -Y_2^{i}\\
-Y_3^{i}& Y_1^{i}\end{array}\right)
 , (\sigma_i^2)^{-1}=\frac{1}{Y_1^{i}Y_4^{i}-Y_2^{i}Y_3^{i}}
 \left( \begin{array}{cc}Y_4^{i}& -Y_2^{i}\\ Y_3^{i}&
 -Y_1^{i}\end{array}\right), \nonumber \\
 (\sigma_i^3)^{-1}&=&\frac{1}{Y_1^{i}Y_4^{i}-Y_2^{i}Y_3^{i}}
 \left( \begin{array}{cc}-Y_3^{i}& Y_1^{i}\\ Y_4^{i}&
 -Y_2^{i}\end{array}\right),
  (\sigma_i^4)^{-1}=\frac{1}{Y_1^{i}Y_4^{i}-Y_2^{i}Y_3^{i}}\left(
\begin{array}{cc}-Y_3^{i}& Y_1^{i}
\\ -Y_4^{i}&
  Y_2^{i}\end{array}\right).
\end{eqnarray}}

The universal expression of the inverse sub-matrix is {\footnotesize
\begin{eqnarray}
(\sigma_i^{\mu})^{-1}&=&\frac{1}{(T_{\mu1}^{i}Y_{1}^{i}
+T_{\mu2}^{i}Y_{2}^{i})(T_{\mu3}^{i}Y_{3}^{i}
 +T_{\mu4}^{i}Y_{4}^{i})-(T_{\mu3}^{i}Y_{1}^{i}
+T_{\mu4}^{i}Y_{2}^{i})(T_{\mu1}^{i}Y_{3}^{i}+T_{\mu2}^{i}Y_{4}^{i})}
\left( \begin{array}{cc}T_{\mu3}^{i}Y_{3}^{i}
 +T_{\mu4}^{i}Y_{4}^{i}
& -T_{\mu3}^{i}Y_{1}^{i}-T_{\mu4}^{i}Y_{2}^{i} \\
-T_{\mu1}^{i}Y_{3}^{i}-T_{\mu2}^{i}Y_{4}^{i} & T_{\mu1}^{i}Y_{1}^{i}
+T_{\mu2}^{i}Y_{2}^{i}\end{array}\right).
\end{eqnarray}}

  Obviously $Y_1^{i}Y_4^{i}\neq Y_2^{i}Y_3^{i}$ is the necessary condition of the
successful teleportation. The existence of the inverse matrix can be
used as a criterion to judge if the teleportation is successful. How
to realize such an inverse transformation is still an open problem.

(1) Using the decomposition matrix to forecast the collapsed quantum
state of the receiver

When the sender Alice measures $|\Psi_T>$ with one of
$|\varphi_{1,5}^{\alpha_1}\varphi_{2,7}^{\alpha_2}\varphi_{3,9}^{\alpha_3}>$,
the receiver Bob can obtain the collapsed quantum states
$(1/\sqrt{2})^3\sigma_{4,6,8}^{\alpha_{1}\alpha_{2}\alpha_{3}}|\varphi_{4,6,8}>$,
which are determined by the decomposition matrix.

In Ref.\cite{ZhengAPS52_2678}, the unknown state and the quantum
channel satisfy $Y_2^{1}=Y_3^{1}=Y_2^{2}=Y_3^{2}=Y_2^{3}=Y_3^{3}=0$,
$|\varphi>_{1,2,3}=X_1|000>+X_2|001>+X_3|010>+X_6|100>$,
$|\varphi>_{4,5}=Y_1^{1}|00>+Y_4^{1}|11>,\
|\varphi>_{6,7}=Y_1^{2}|00>+Y_4^{2}|11>,\
|\varphi>_{8,9}=Y_1^{3}|00>+Y_4^{3}|11>$. The sub-matrixes of the
quantum
channel are {\footnotesize $\sigma_4^1=\left( \begin{array}{cc}Y_1^{1}& 0 \\
0 & Y_4^{1}\end{array}\right),\ \
 \sigma_4^2=\left( \begin{array}{cc}Y_1^{1}& 0 \\ 0 &
 -Y_4^{1}\end{array}\right)
 , \sigma_4^3=\left( \begin{array}{cc}0 & Y_1^{1}\\ Y_4^{1}&
 0\end{array}\right)
  , \sigma_4^4=\left( \begin{array}{cc}0 & -Y_1^{1}\\ Y_4^{1}&
  0\end{array}\right)$},
  {\footnotesize $\sigma_6^1=\left( \begin{array}{cc}Y_1^{2}& 0 \\
0 &
 Y_4^{3}\end{array}\right)
 , \sigma_6^2=\left( \begin{array}{cc}Y_1^{2}& 0 \\ 0 &
 -Y_4^{3}\end{array}\right)
 , \sigma_6^3=\left( \begin{array}{cc}0 & Y_1^{2}\\ Y_4^{2}&
 0\end{array}\right)
  , \sigma_6^4=\left( \begin{array}{cc}0 & -Y_1^{2}\\ Y_4^{2}&
  0\end{array}\right)$},
  {\footnotesize $\sigma_8^1=\left( \begin{array}{cc}Y_1^{3}& 0 \\
0 &
 Y_4^{3}\end{array}\right)
 , \sigma_8^2=\left( \begin{array}{cc}Y_1^{3}& 0 \\ 0 &
 -Y_4^{3}\end{array}\right)
 , \sigma_8^3=\left( \begin{array}{cc}0 & Y_1^{3}\\ Y_4^{3}&
 0\end{array}\right)
  , \sigma_8^4=\left( \begin{array}{cc}0 & -Y_1^{3}\\ Y_4^{3}&
  0\end{array}\right)$}. From Eq.(8) we know that
  $(1/\sqrt{2})^3\sigma_{4,6,8}^{\alpha_{1}\alpha_{2}\alpha_{3}}|\varphi_{4,6,8}>$
  is the collapsed state that Bob can manipulate, taking the sub-matrixes we get the
  same results as in Ref.\cite{ZhengAPS52_2678}.

 (2) Using the decomposition matrix to determine the manipulation of the receiver

For the maximally entangled channel ($Y_1^{i}=Y_4^{i}=1/\sqrt{2},
  Y_2^{i}=Y_3^{i}=0$), $\sigma^1=\sigma_0/\sqrt{2},\
  \sigma^2=\sigma_z/\sqrt{2},\ \sigma^3=\sigma_x/\sqrt{2}$ and $
  \sigma^4=-(i\sigma_y)/\sqrt{2}$, the corresponding inverse matrix
  are $(\sigma^1)^{-1}=\sqrt{2}\sigma_0,\
  (\sigma^2)^{-1}=\sqrt{2}\sigma_z,\ (\sigma^3)^{-1}=\sqrt{2}\sigma_x$ and $
  (\sigma^4)^{-1}=\sqrt{2}(i\sigma_y)$, where $\sigma_{x,y,z}$ is the Pauli
  matrix and $\sigma_0$ is the identity matrix. Here
  $(\sigma^{\mu})^{-1}$ is the transformation matrix that Bob can manipulate.

  (3) Using the decomposition matrix to calculate the probability of successful teleportation

The probability of successful teleportation is determined by the
parameters of the quantum channel, while the answer of the GGNM
problem is very complex and tedious, we will discuss them later.

\section{5. Conclusions}

In this paper, we propose a method of the Bell bases decomposition
to teleport an unknown N-qubit state through a non-maximally
entangled quantum channel, and give the general expression of the
decomposition matrix for N-qubit case. In theory, we solved the
three key problems in GGNM teleportation. Using the decomposition
matrix, Bob can easily obtain the collapsed state, the inverse
matrix of the decomposition matrix is just the transformation matrix
that Bob can manipulate. The sub-matrixes
($\prod_{\alpha_{i}=1}^N\otimes\sigma^{\alpha_{i}}$) of the
decomposition matrix is determined by the parameters of the quantum
channel, the inverse matrix of the sub-matrix determines the
transformation matrix that Bob can manipulate and its parameters
determine the probability of the successful transformation, the
existence of the inverse matrix is a criterion that can help us to
judge if the teleportation is successful.

There still exists some open problems, for example, how to realize
the transformation of the inverse matrixes and how to generalize
this method to the quantum channel with smaller assistant qubits.

 \vspace{0.5cm} We would like to thank Pro. X. W. Zha for his warmhearted discussions.
  This work is supported by the NSF of
China under grant 10547008 and the NSF of Shaanxi Province under
grant 2004A15.

\end{document}